\documentclass[aps,pre,twocolumn,showpacs,superscriptaddress,groupedaddress]{revtex4-2}
\pdfoutput=1
\usepackage{hyperref}
\usepackage{epsfig}
\usepackage{amsmath}
\usepackage{graphicx}
\usepackage{wrapfig}
\usepackage{dcolumn}
\usepackage{bm}
\usepackage{amssymb}
\usepackage{titlesec}
\usepackage{mathtools}
\usepackage{relsize}
\usepackage{cleveref}
\usepackage{calligra}
\usepackage{bigints}
\usepackage{ulem}
\usepackage{scalerel}
\usepackage{float}
\usepackage[usenames,dvipsnames]{color}

\begin{document}
\title{Outbreak-size distributions under fluctuating rates}
\author{Jason Hindes$^{1}$, Luis Mier-y-Teran-Romero$^{\;2}$, Ira B. Schwartz$^{1}$ and Michael Assaf$^{\;3}$}
\affiliation{$^{1}$U.S. Naval Research Laboratory, Washington, DC 20375, USA}
\affiliation{$^{2}$Leidos, Reston, VA 20190, USA}
\affiliation{$^{3}$Racah Institute of Physics, Hebrew University of Jerusalem, Jerusalem 91904, Israel}

\begin{abstract}
We study the effect of noisy infection (contact) and recovery rates on the distribution of outbreak sizes in the stochastic SIR model. The rates are modeled as Ornstein-Uhlenbeck processes with finite correlation time and variance, which we illustrate using outbreak data from the RSV 2019-2020 season in the US. In the limit of large populations, we find analytical solutions for the outbreak-size distribution in the long-correlated (adiabatic) and short-correlated (white) noise regimes, and demonstrate that the distribution can be highly skewed with significant probabilities for large fluctuations away from mean-field theory. Furthermore, we assess the relative contribution of demographic and reaction-rate noise on the outbreak-size variance, and show that demographic noise becomes irrelevant in the presence of slowly varying reaction-rate noise but persists for large system sizes if the noise is fast. Finally, we show that the crossover to the white-noise regime typically occurs for correlation times that are on the same order as the characteristic recovery time in the model.  
\end{abstract}

\maketitle

\section{\label{sec:Intro}INTRODUCTION}
Epidemic models are useful for understanding the spreading dynamics of general contagious processes and effectively describe a wide variety of phenomena from spreading diseases, to rumors, fads, panics, computer viruses, and even election dynamics\cite{keeling:infectious_diseases,AnderssonBook,RevModPhys.87.925,rodrigues2016application,billings2002unified,RevModPhys.80.1275,hindes2019degree,10.1137/19M1306658}. In addition, epidemic models are practically useful, since epidemiologists rely on models to quantify risks of local epidemic outbreaks of various sizes and formulate optimal control strategies for many diseases including Sars-Cov-2, Ebola, and Dengue\cite{ModelingCOVID-19, Ghaemi2022, RodriguezBarraquer2014,Catching2020.08.12.20173047,hindes2022outbreak,10.1371/journal.pone.0244706}. Within a given population, outbreak dynamics are typically described in terms of compartmental models\cite{keeling:infectious_diseases,MathematicalepidemiologyPastPresenFuture,rodrigues2016application}. For example, starting from some small initial infection, over time, individuals in a population make stochastic transitions between some number of discrete disease states (susceptible, exposed, infectious, etc.) based on prescribed probabilities for a particular population
and disease\cite{ModelingCOVID-19,Ray2020.08.19.20177493,Catching2020.08.12.20173047,doi:10.1146/annurev-statistics-061120-034438,SIRSi,miller2019distribution,Aron1984}. In the limit of large populations and non-fluctuating parameters the stochastic dynamics approach deterministic (mean-field) differential equations for the expected fraction of a population in each state\cite{keeling:infectious_diseases,MathematicalepidemiologyPastPresenFuture,rodrigues2016application,StochasticEpidemicModels}. 

Yet, for real finite populations with time-fluctuating parameters outbreak dynamics have a wide range of outcomes for each initial condition, which are not predicted by mean-field models~\cite{doi:10.1146/annurev-statistics-061120-034438,keeling:infectious_diseases,StochasticEpidemicModels,doi:10.1098/rspa.2012.0436,ball_1986,ball_clancy_1993,miller2019distribution,Dykman1994}. For instance, recently we developed a theoretical approach that allows for calculating the distribution of outbreak sizes in well-mixed populations under demographic noise. This approach provided a closed-form expression for the asymptotic outbreak distribution in the Susceptible-
Infected-Recovered (SIR) model and more general SIR model extensions with fixed population sizes ($N$) and static infection/recovery rates\cite{hindes2022outbreak}. However, many data analyses have shown that for a multitude of diseases, best-fit epidemic model parameters can fluctuate significantly in time\cite{10.1371/journal.pone.0236464,2020PNAS..117.5067K,EbolaR0inTime,SriLankaR0inTime,QuatarR0inTime,RoleOfTimeVaryingReportingRate,10.1093/biostatistics/kxs052,COVIDstudyWithStochastic}. For instance, by measuring the relative changes in reported disease incidence and hospitalization, one can compute an effective infectious contact rate between individuals in a population over time. Doing so one often finds fluctuating and/or periodic rates in general\cite{London1973,Yorke1979,Fine1980,Grenfell2002,Schwartz1992,keeling:infectious_diseases,RoleOfTimeVaryingReportingRate,10.1093/biostatistics/kxs052}, which in the case of human epidemics correlate with more general social contact rates\cite{TimeTrendsInSocialContacts}.
For instance, techniques for extracting time-dependent parameters have been applied to the recent COVID-19 pandemic as well\cite{Chen_2020,Setianto2023,RoleOfTimeVaryingReportingRate,COVIDstudyWithStochastic}, in order to account for fluctuations in contact rates, rendering the usual SIR class of forecasting models time dependent. 
In addition, here we give an another example based on 2019-2020 hospitalization data of the respiratory syncytial virus (RSV) season in the US\cite{CDCdata}, and find the data effectively parameterized in terms of two general metrics for quantifying temporal variations about a mean: the infection rate's standard deviation and correlation time.

Despite the theoretical importance of understanding noise effects in canonical non-equilibrium epidemic models, as well as the practical importance for quantifying uncertainty in real epidemics, a general analytical approach for describing small and large fluctuations in outbreak dynamics due to parameter fluctuations is still lacking. Here we develop such an approach within the context of the SIR model with noisy reaction rates with finite variances and correlation times. We motivate our use of these standard noise characteristics by extracting infectious contact rate fluctuations in RSV outbreak data from the U.S. in 2019-2020 using a Bayesian model inference. In terms of general model analysis, we focus on the outbreak-size distribution and quantify the probabilities for outbreaks that differ from the mean-field predictions. In particular, we calculate the distribution  in the limit of adiabatic and white noises, and demonstrate several important properties including: the skewness of the outbreak distribution toward unusually small outbreaks, and the existence of optimal values of the basic reproductive number that maximize the outbreak variance. We also study the cross-over of the outbreak distribution with finite population size and noise-correlation time and analyze when the limiting theories of demographic, adiabatic and white reaction-rate noise apply. 

\section{\label{sec:SIRModel}SIR MODEL WITH REACTION-RATE NOISE}
We are interested in outbreak dynamics in which the epidemic wave is fast compared to both demographic and re-infection time scales; the latter denotes the possibility for individuals to be infected multiple timescales\footnote{If re-infection happens on short timescales, then models such as SIRS may be more appropriate\cite{keeling:infectious_diseases}}. The canonical epidemic model for this regime is the SIR model\cite{keeling:infectious_diseases,AnderssonBook}, in which individuals are either susceptible (capable of getting infected), infected, or recovered (or removed$/$deceased), and can make transitions between these states through two basic processes: infection and recovery. Denoting the total number of susceptibles $S$, infected $I$, and recovered $R$ in a population of fixed size $N$, the probability per unit time that the number of susceptibles decreases by one and the number of infected increases by one is $\beta SI\!/\!N$ (for a well-mixed population), where $\beta$ is the infectious contact rate\cite{keeling:infectious_diseases,AnderssonBook,rodrigues2016application}. Similarly, the probability per unit time that the number of infected decreases by one is $\gamma I$, where $\gamma$ is the recovery rate\cite{keeling:infectious_diseases,AnderssonBook,rodrigues2016application}. 
As a result, the deterministic rate equations in the limit of large $N$ describing the mean \textit{densities} of susceptibles, infected and recovered, $x_s=S/N$, $x_i=I/N$ and $x_r=R/N$, respectively read:
\begin{equation}
\label{MFdynamics}
\dot{x}_s=-\beta x_s x_i,\quad\dot{x}_i=\beta x_s x_i-\gamma x_i,\quad \dot{x}_r=\gamma x_i,
\end{equation}
where $x_s+x_i+x_r=1$. Starting from a small initial infection density, $x_{i}(t\!=\!0)\!\ll\!1$, the final fraction of susceptibles in Eqs.(\ref{MFdynamics}) $\overline{x_s^*}\equiv x_0$ satisfies $x_0=e^{-R_0(1-x_0)}$, where $R_0=\beta/\gamma$ is the basic reproduction number\cite{keeling:infectious_diseases,AnderssonBook}. Hence, in the mean-field theory the total fraction of the population infected over the whole epidemic wave is $\overline{x_r^*}=1-x_0$, 
\begin{equation}
\label{MFSteadyState}
\overline{x_r^*}=1-e^{-R_0 \overline{x_r^*}}. 
\end{equation} 
Note that if $R_{0}\leq 1$ in Eq.(\ref{MFSteadyState}) then $\overline{x_r^*}=0$, giving us the usual condition $R_{0}\!=\!1$ as the epidemic threshold.  

As noted in Sec.\ref{sec:Intro}, in many cases the parameters for the SIR model are time fluctuating. As a simple model, we allow the infection and recovery rates to be generated by independent Ornstein-Uhlenbeck (OU) processes with some correlation times and variances. For simplicity, we assume  the correlation times are identical for both rates and equal $\tau$,  while the noise variances are $\sigma_{\beta}^{2}$ and $\sigma_{\gamma}^{2}$ for the infection and recovery rates, respectively. Thus, we write: $\beta(t)=\beta_0(1+\xi_{\beta}(t))$ and $\gamma(t)=\gamma_0(1+\xi_{\gamma}(t))$, and  
augment Eq.(\ref{MFdynamics}) into the stochastic system  
\begin{eqnarray}\label{Langeqs}
\dot{x}_s&=&-\beta_0(1+\xi_{\beta}) x_s x_i,\nonumber\\
\dot{x}_i&=&\beta_0(1+\xi_{\beta}) x_s x_i-\gamma_0(1+\xi_{\gamma})x_{i},\nonumber\\
\dot{\xi}_{\beta}&=&-\frac{\xi_{\beta}}{\tau}+\sqrt{\frac{2}{\tau}}\Big(\frac{\sigma_{\beta}}{\beta_{0}}\Big)\eta_{\beta}(t),\nonumber\\
\dot{\xi}_{\gamma}&=&-\frac{\xi_{\gamma}}{\tau}+\sqrt{\frac{2}{\tau}}\Big(\frac{\sigma_{\gamma}}{\gamma_{0}}\Big)\eta_{\gamma}(t).
\end{eqnarray}
Here, $\eta_{\beta}$ and $\eta_{\gamma}$ are Gaussian white noises, while $\xi_{\beta}$ and $\xi_{\gamma}$
are OU processes. Note that by construction, $\beta(t)$ and $\gamma(t)$ are assumed to be wide-sense stationary Gaussian processes with $\left<\beta\right>\!=\!\beta_{0}$, $\left<\gamma\right>\!=\!\gamma_{0}$, 
$\left<(\beta(t)-\beta_{0})(\beta(t+\Delta)-\beta_{0})\right>\!=\!\sigma_{\beta}^{2}e^{-\Delta/\tau}$, and 
$\left<(\gamma(t)-\gamma_{0})(\gamma(t+\Delta)-\gamma_{0})\right>\!=\!\sigma_{\gamma}^{2}e^{-\Delta/\tau}$, where $\left<\cdot\right>$ denotes the expectation operator. In general, one can simulate the system of equations~(\ref{Langeqs}) and find the final outbreak-size distribution for fluctuating rates with any magnitude and correlation time.

\begin{figure}[H]
\center{\includegraphics[scale=0.217]{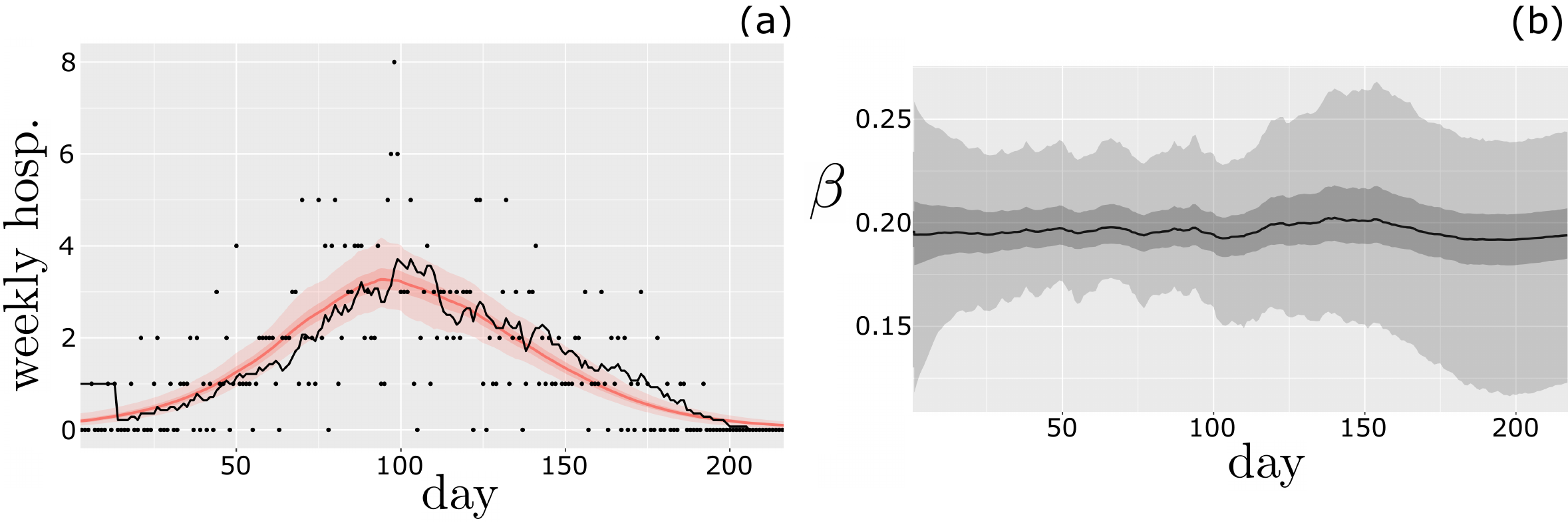}}
\vspace{-7mm}
\caption{RSV model inference. (a) Weekly RSV hospitalizations (black dots) and 2-week rolling average (black line) from the 2019-2020 season in the United States\cite{CDCdata}. Results from the Bayesian inference model are overlaid with the data (median: red line, shaded bands represent the inter-quartile range and the 95\% credible intervals). (b) Inferred infectious contact rate obeying a time-discretized version of the OU process (median: black line, shaded bands represent the inter-quartile range and the 95\% credible intervals).}
\label{fig0}
\end{figure}

\subsection{\label{sec:RSVModelFit} RSV model fit}
Finite correlation time and variance are general physical metrics that quantify temporal fluctuations around a mean -- the sort of temporal variation observed in many epidemic data analyses\cite{10.1371/journal.pone.0236464,2020PNAS..117.5067K,EbolaR0inTime,SriLankaR0inTime,QuatarR0inTime,RoleOfTimeVaryingReportingRate,10.1093/biostatistics/kxs052,COVIDstudyWithStochastic,London1973,Yorke1979,Fine1980,Grenfell2002,Schwartz1992,keeling:infectious_diseases,RoleOfTimeVaryingReportingRate,10.1093/biostatistics/kxs052,Chen_2020,Setianto2023}. We can further motivate our study of the SIR model with temporally fluctuating reaction rates by extracting such noise characteristics from data on the 2019-2020 respiratory syncytial virus (RSV) season in the U.S. 

We perform a parameter inference from RSV hospitalization data assuming a discretized version of Eqs.(\ref{Langeqs}) with daily time steps and a fixed recovery rate $\gamma\!=\!\gamma_{0}\!=\!1/7\;\text{days}^{-1}$ \cite{RSVseasonality}. We use the well-known platform {\tt Stan} via the {\tt R} package {\tt rstan} \cite{Rstan,STANexampleCOVID} to do the numerical Bayesian inference by tying the dynamical model to the number of recorded daily hospitalizations, as obtained from the CDC\cite{CDCdata}. The parameters for the inference are: $\beta_{0}$, the inverse correlation time $\alpha$, $\sigma_{\beta}$, the hospitalization rate, and initial conditions for the SIR\footnote{The Bayesian framework estimates the initial state for the SIR model that is used for prediction in a single year. The state includes the fraction of the population recovered from previous years. Since, the model does not include birth, death, re-infection, etc. it is not appropriate for multiple-year predictions\cite{RSVseasonality2}, but is reasonable for a single season.}; output examples  are shown  in Fig.\ref{fig0}. In panel (a) we plot the daily hospitalization numbers and compare to the median prediction of the model (in red) along with its credible intervals. A similar plot is shown in panel (b) for the daily infectious contact rate, which drives the predictions for (a). Further details are given in App.\ref{sec:RSV_DataAnalysis} \cite{Code}.

Our inference uncovers significant temporal fluctuations in the most-likely RSV infectious contact rate. A summary of the output that is relevant for our analysis includes: $\hat{R}_{0}\!=\!1.37$ in $[1.32,1.44]$, $\hat{\alpha}\!=\!0.11\;\text{days}^{-1}$ in $[0.045,0.20]\;\text{days}^{-1}$, and 
$\hat{\sigma}_{\beta}\!=\!0.026\;\text{days}^{-1}$ in $[0.014,0.040]\;\text{days}^{-1}$, where $\hat{\;\;}$ denotes the median within a quartile range specified by the square brackets. From these we observe a fairly tight value of the inferred time-averaged $R_{0}$, but with substantial temporal fluctuations between $10-20\%$. On the other hand, the correlation time estimate $\alpha^{-1}$ is quite broad ranging from $5-20$ days. Nevertheless, note that such time scales are significantly smaller than seasonal effects, which are expected to occur on time scales on the order of a year. The noise-inference, therefore, quantifies temporal fluctuations distinct from seasonality\cite{RSVseasonality2}. We can situate the inferred noise characteristics of the RSV season within the results of our analytical theory, see Sec{\ref{sec:Crossover}}. We begin by analyzing outbreak statistics driven by the fluctuations in Eqs.(\ref{Langeqs}).   

\section{\label{sec:LimitAdiabatic} LIMIT OF ADIABATIC NOISE}
 In order to gain insight on the outbreak distributions generated from the general Eqs.(\ref{Langeqs}) and temporal fluctuations of the sort we inferred from RSV data, we first consider limiting regimes. We start with the limit of adiabatic noise, $\tau\gg 1$. Here, the underlying assumption is that, during the epidemic wave, the rates do not change appreciably, and hence the problem reduces to that of quenched noise on system (\ref{MFdynamics}). For simplicity and illustration of the adiabatic limit, here we deal with the case where only $\beta$ varies and $\gamma$ is constant, such that $\sigma_\gamma\!=\!0$. In order to simplify the equations, we take $\gamma_0\!=\!1$, which merely specifies the time units and results in $R_{0}\!=\!\beta$. 

To find the distribution of the final outbreak size $P(x_r^*)$, we have to compute the following integral:
\begin{equation}\label{adia1}
P(x_r^*)=\int_{-\infty}^{\infty}P(x_r^*|\beta)P(\beta)d\beta. 
\end{equation}
The conditional probability $P(x_r^*|\beta)$ is a dirac delta function around the mean-field value of the outbreak at $\beta$. Namely, $P(x_r^*|\beta)=\delta(x_r^*-\overline{x_r^*})$, where the mean-field final outbreak fraction $\overline{x_r^*}$ satisfies Eq.(\ref{MFSteadyState}). Taking a Gaussian distribution for $P(\beta)$ with mean $\beta_0$ and standard deviation $\sigma_{\beta}$, Eq.~(\ref{adia1}) becomes
\begin{equation}\label{adia2}
P(x_r^*)=\frac{1}{\sqrt{2\pi\sigma_{\beta}^2}}\int_{1}^{\infty}\delta(x_r^*-\overline{x_r^*}(\beta))e^{-\frac{(\beta-\beta_0)^2}{2\sigma_{\beta}^2}}d\beta.
\end{equation}
We point out that in order for the SIR model to make physical sense $\beta\!\geq\!0$. Therefore, when plugging in an unrestricted Gaussian in Eq.(\ref{adia2}), $\sigma_{\beta}$ cannot be too large~\footnote{we must assume that e.g., $\sigma_{\beta}/\beta_{0}\lesssim 1/3$ so that over $99\%$ of the distribution is in the physical domain.}. Otherwise other distributions, e.g. that vanish at $\beta=0$ can be used instead; yet this does not change the results qualitatively. 
We also note that since $\overline{x_r^*}(\beta)$ vanishes for $\beta\leq 1$, the lower boundary in the integral in Eq.~(\ref{adia2}) can be taken to be $1$ without affecting the distribution. 

Changing variables from $\beta$ to $x_r^*$, and using the fact that $d\beta/dx_r^*=[x_r^*+(1-x_r^*)\ln(1-x_r^*)]/[(1-x_r^*)(x_r^*)^2]$, we can explicitly perform the integration by plugging instead of $\beta$, $-\log(1-x_r^*)/x_r^*$, which is the solution of $x_r^*=\overline{x_r^*}(\beta)$. As a result, Eq.~(\ref{adia2}) reduces to
\begin{equation}\label{adia3}
P(x_r^*)=\frac{d\beta}{dx_r^*}\exp\!\left[-\frac{\left((1/x_r^*)\ln(1-x_r^*)+\beta_0\right)^2}{2\sigma_{\beta}^2}\right].
\end{equation}
From Eq.(\ref{adia3}) we can derive, e.g., the typical fluctuations around the mean-field given by $\sigma_{\text{a}}$ -- the standard deviation associated with the adiabatic outbreak PDF. In particular, in the limit of small $\sigma_{\beta}$, the variance becomes
\begin{equation}\label{SigmaAdi}\hspace{-2mm}
\sigma_{\text{a}}\!=\!\sigma_{\beta}\!\left.\!\frac{dx_r^*(\beta)}{d\beta}\!\right|_{\beta=\beta_0}\!=\frac{\sigma_{\beta}x_0(1-x_0)}{1-R_0 x_0}.
\end{equation}
Note that for adiabatic noise we can repeat our calculation of the outbreak-size distribution and variance for any quenched distribution of $\beta$ (or $\gamma$), and not just Gaussians, which may better reflect the data in a given application. 
\begin{figure}[t]
\center{\includegraphics[scale=0.217]{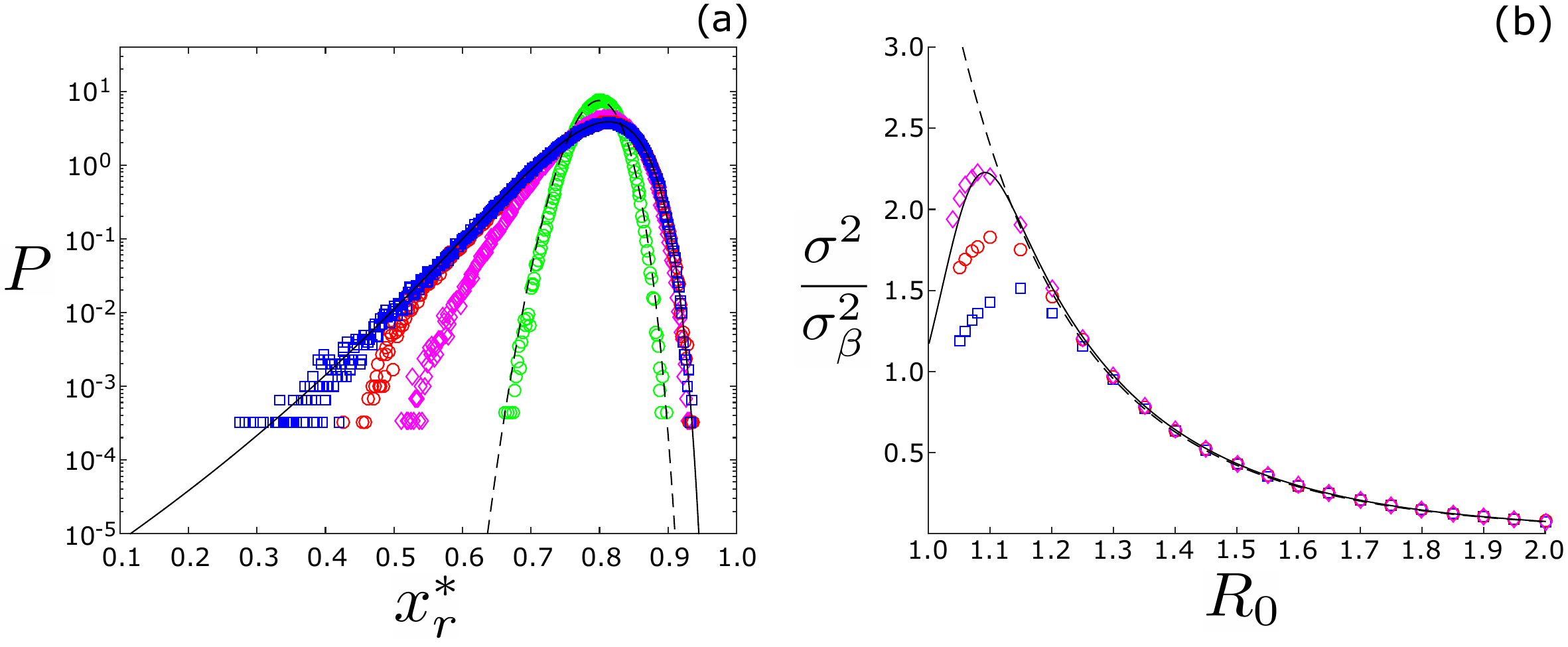}}
\vspace{-7mm}\caption{Outbreak statistics for adiabatic noise. (a) Simulated PDFs of the final outbreak fraction $x_r^*$ from Eqs.(\ref{Langeqs}), in the case of time-correlated reaction-rate noise for $\tau=10^0$-$10^3$ (from narrowest to widest) with $\sigma_{\beta}=0.1\beta_{0}$ and $\beta_{0}\!=\!2$. The solid and dashed lines denote the adiabatic and white-noise predictions, respectively. (b) Variance of the outbreak PDF (normalized by infection noise) versus $R_0\!=\!\beta_{0}/\gamma_{0}$ for $\tau=10^3$-$10^5$ (bottom to top) and $\sigma_{\beta}=0.04\beta_{0}$. The variance of the PDF~(\ref{adia3}) is shown in solid black, while the small-noise limit Eq.(\ref{SigmaAdi}) is shown with a dashed line. For both panels $\gamma_{0}\!=\!1$.}
\label{fig1}
\end{figure}

Next, we can plot the probability distribution function (PDF) for adiabatic infection-rate noise and explore its qualitative features. An example prediction is shown in Fig.\ref{fig1} (a) with a solid line for fixed values of $\beta_{0}\!=\!2$ and $\sigma_{\beta}=0.1\beta_{0}$. The solution from Eq.(\ref{adia3}) can be compared to stochastic simulations of Eqs.(\ref{Langeqs}) for large $\tau$. Note that the agreement with simulations is quite good. Qualitatively, one of the most important features that we observe in the PDFs is the high degree of skewness toward small outbreaks. We can get a quantitative measure of this skewness by examining the exponent of Eq.(\ref{adia3}), called the \textit{action} (for reasons explained in Sec.\ref{sec:LimitWhite}), for two limiting values of the outbreak fraction: $x_{r}^{*}\!=\!0$ and $x_{r}^{*}\!=\!1$, i.e., small and large outbreaks. Indeed,
the PDF [Eq.(\ref{adia3})] can be described effectively as $P\!\sim\!\exp[-S/\sigma_{\beta}^{2}]$, where $S\!=\!(\ln(1-x_{r}^{*})/x_r^*+R_{0})^{2}/2$. 
When $x_r^{*}\!\rightarrow\!0$ the action remains finite, i.e, $S\!\rightarrow\!(R_{0}-1)^{2}/2$. On the other hand, when $x_r^{*}\!\rightarrow\!1$, $S\!\rightarrow\!\infty$. Hence, minimally small outbreaks 
occur with finite probability for finite $R_{0}$, while maximally large outbreaks can never occur when the reaction-rate noise is finite, which is why the outbreak distribution's tails are skewed toward small outbreaks. 

In addition to the PDFs, we can examine the variance of the outbreak PDF for adiabatic noise as a function of $R_{0}$. Examples can be seen in Fig.\ref{fig1}(b), where we plot simulated outbreak variances for three large values of $\tau$ with $\gamma\!=\!1$. Here another interesting qualitative feature emerges: the existence of a maximum in the outbreak variance for some value of $R_{0}$.   
On the one hand, as $\sigma_{\beta}\to 0$, the maximum approaches $R_{0}\!=\!1$. On the other hand as $\sigma_{\beta}$ increases, the maximum variance occurs for an $R_{0}$ that is an increasing function of $\sigma_{\beta}$. For example, in Fig.\ref{fig1} (b) we observed a maximum near $R_0\!=\!1.1$. However, the saddle-point equation for the maximum variance in the adiabatic limit cannot be solved analytically.  

In general, we observe good agreement with the predicted variance of Eq.(\ref{adia3}) (solid line) and the small-noise limit Eq.(\ref{SigmaAdi}) (dashed line), including the existence of a maximum, which the former captures. Yet, as $R_{0}\!\rightarrow\!1$, eventually all  the simulation results have discrepancy with both adiabatic predictions. The reason is, as we approach the epidemic threshold, the SIR dynamics slow down, meaning that even a large $\tau$  may not be ``slow" with respect to the underlying process.

\section{\label{sec:LimitWhite} LIMIT OF WHITE NOISE}
So far we have assumed that the dynamics of the noise is slow compared to the dynamics of Eqs.(\ref{MFdynamics}), but what happens if it is fast? In this latter limit, $\tau\to 0$, instead of Eqs.~(\ref{Langeqs}) we can write 
\begin{eqnarray}\label{white}
\dot{x}_s&=&-\beta_0(1+\sigma_{1}\zeta_{\beta}(t)) x_s x_i,\nonumber\\
\dot{x}_i&=&\beta_0(1+\sigma_{1}\zeta_{\beta}(t)) x_s x_i-\gamma_0(1+\sigma_{2}\zeta_{\gamma}(t))x_i, 
\end{eqnarray}
which we denote as the white reaction-rate limit. Here, $\zeta_{\beta}$ and $\zeta_{\gamma}$ are white Gaussian noises. In order to coincide with Eqs.~(\ref{Langeqs}) as $\tau\!\to\! 0$, one must demand that 
\begin{eqnarray}\label{whiteNoiseAmplitudes}
\sigma_{1}\!=\!\sqrt{2\tau}\sigma_{\beta}/\beta_{0},\quad
\sigma_{2}\!=\!\sqrt{2\tau}\sigma_{\gamma}/\gamma_{0}.
\end{eqnarray}

\begin{figure}[t]
\center{\includegraphics[scale=0.202]{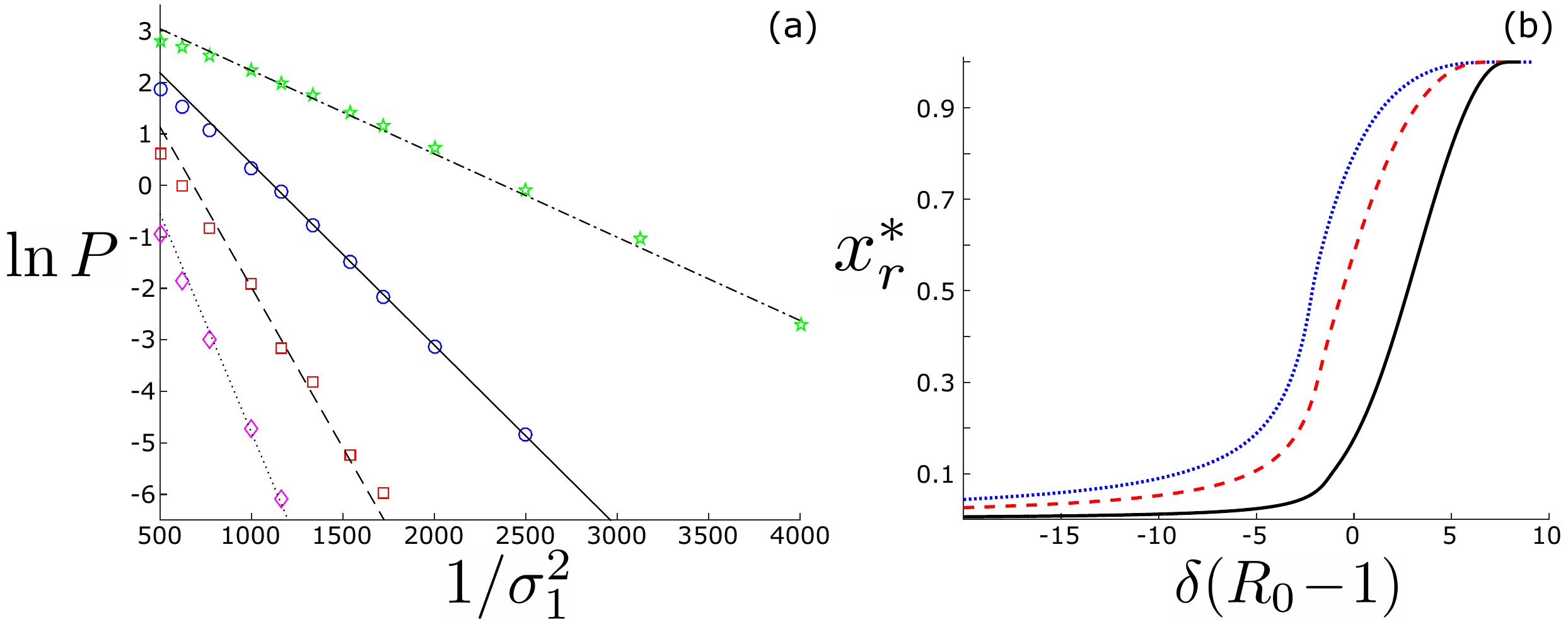}}
\vspace{-3mm}\caption{Scaling of the outbreak-size distribution. (a) The natural log of simulated PDF values for $x_{r}^{*}\!=\!0.785$ (green stars), $0.780$ (blue circles), $0.775$ (red squares), and $0.770$ (magenta diamonds) from Eqs.(\ref{Langeqs}) versus the reaction-rate noise variance. The slopes are predictions from the white-noise theory. Other model parameters are $\beta_{0}\!=\!2$, $\gamma_{0}\!=\!1$, $\tau\!=\!0.1$, and $f\!=\!0$. (b) The final outbreak size plotted as a function of the infected momentum initial condition multiplied by $R_{0}\!-\!1$ for: $R_{0}\!=\!1.1$ (solid black), $R_{0}\!=\!1.r$ (dashed red), and $R_{0}\!=\!2$ (dotted blue). Other model parameters are $\gamma_{0}\!=\!1$ and $f\!=\!0.65$.}
\label{fig2}
\end{figure}

To analyze the outbreak-size PDF given Eq.~(\ref{white}),
we follow the approach  in\cite{hindes2022outbreak}, and  construct the equivalent Fokker-Planck equation for the probability to observe densities $x_s$ and $x_i$ at time $t$ (assuming It\^{o} calculus):
\begin{eqnarray}
\label{FPwn}
\dfrac{\partial P}{\partial t} =& -\dfrac{\partial}{\partial x_{s}}\Big[\!\!-\!\beta_{0}x_{s}x_{i} P\Big] -\dfrac{\partial}{\partial x_{i}}\Big[(\beta_{0}x_{s}x_{i}-\gamma_{0}x_{i})P\Big] \;+ \nonumber\\
&\!\!\!\!\!\!\!\!\!\!\Big(\dfrac{\partial^{2}}{\partial x_{s}^{2}}+\dfrac{\partial^{2}}{\partial x_{i}^{2}}-2\dfrac{\partial}{\partial x_{s}}\dfrac{\partial}{\partial x_{i}}\Big)\Big[\frac{1}{2}\beta_{0}^{2}x_{s}^{2}x_{i}^{2}\sigma_{1}^{2}P\Big] \;+ \nonumber\\
&\!\!\!\!\!\!\!\!\!\!\!\!\!\!\!\!\!\!\!\!\!\!\!\!\!\!\!\!\!\!\!\!\!\!\!\!\!\!\!\!\!\!\!\!\!\!\!\!\!\!\!\!\!\!\!\!\!\!\!\!\!\!\!\!\!\!\!\!\!\!\!\!\!\!\dfrac{\partial^{2}}{\partial x_{i}^{2}}\Big[\frac{1}{2}\gamma_{0}x_{i}^{2}\sigma_{2}^{2}P\Big].  
\end{eqnarray}
To simplify notation, henceforth, we will assume that $\sigma_{2}^2\!=\!f\sigma_{1}^2$, with $f\!>\!0$ for simplicity, and again rescale time $t\to \gamma_0 t$, so that $\beta_0$ is replaced by the basic reproduction number, $R_0=\beta_0/\gamma_0$. Next, we employ the WKB approximation $P(x_s,x_i)\!\sim\!\exp\!\left[-S(x_s,x_i)/\sigma_{1}^2\right]$, which is the expected scaling-form 
for solutions to Eq.(\ref{FPwn}) in the limit of small noise and large deviations\cite{dykman1990large,freidlin2012random,Assaf_2017,doi:10.1137/17M1142028}, and which we observe in simulations of Eqs.(\ref{Langeqs}). Figure \ref{fig2} shows several examples of the expected scaling with noise-variance for different values of the final outbreak size. Indeed, the logarithm of the probability  tends to straight lines as $1/\sigma_{1}^2$ is varied, with slopes that change with the outbreak size. Using this insight, we substitute the exponential ansatz into Eq.(\ref{FPwn}) and arrive at a Hamilton-Jacobi equation, $\partial{S}/\partial{t}\!+\!H\!=\!0$, in the leading order in $\sigma_{1}\ll 1$, with 
\begin{equation}
\label{Hamiltonian}
H=x_i[p_i(R_0 x_s-1)-R_0 x_s  p_s]+\frac{1}{2}x_i^2[R_0^2 x_s^2 (p_s-p_i)^2+f p_i^2]. 
\end{equation}
In this formalism, $H$ is called the Hamiltonian, $S$ is the action, while  $p_s\!=\!\partial_{x_s}S$ and $p_i\!=\!\partial_{x_i}S$ are the conjugate {\it momenta}, just as in analytical mechanics\cite{dykman1990large,Assaf_2017,doi:10.1137/17M1142028}. 

In order to compute probabilities for different outbreak sizes, we need to know the action $S$, which means calculating the integrals $S\!=\!\int p_sdx_s + \int p_idx_i-\int Hdt$ given the dynamics of Hamilton's equations: 
$\dot{x}_s\!=\!\partial_{p_s}H$, $\dot{x}_i\!=\!\partial_{p_i}H$, $\dot{p}_s\!=\!-\partial_{x_s}H$, and $\dot{p}_i\!=\!-\partial_{x_i}H$. 
We can simplify the action computation by noting that, first, since we are interested in outbreaks that emerge from initially small levels of infection $x_{i}(t\!=\!0)\!\rightarrow\!0$ 
the ``energy" of outbreaks is zero,  $H\!=\!0$. As the Hamiltonian~(\ref{Hamiltonian}) has no explicit time dependence, it is a constant of motion, namely $H(t)\!=\!0$. Second, we can rewrite the Hamiltonian as $H\!=\!p_{s}\dot{x}_{s}+p_{i}\dot{x}_{i}\!-\!(1/2)R_{0}^{2}x_{s}^{2}x_{i}^{2}\big[p_{s}-p_{i}\big]^{2}\!-\!(1/2)fx_{i}^{2}p_{i}^{2}$, using $\dot{x}_{s}$ and $\dot{x}_{i}$. Third, by substituting the zero-energy condition into $\dot{p}_{i}$, we obtain that $\dot{p}_{i}=-(1/2)R_{0}^{2}x_{s}^{2}x_{i}\big[p_{s}-p_{i}\big]^{2} -(1/2)fx_{i}p_{i}^{2}$. As a result, the Hamiltonian~(\ref{Hamiltonian}) simplifies to:
$H(t)=p_{s}\dot{x}_{s}+(d/dt)(x_{i}p_{i})$.
Integrating both sides of this equation with respect to time over the full course of an outbreak, yields $0=\int\!p_{s}dx_{s} +x_{i}(t\rightarrow\infty)p_{i}(t\rightarrow\infty)-x_{i}(t\!=\!0)p_{i}(t\!=\!0)$.
As the fraction of the population infected goes to zero both at short and long times (assuming no reinfection), we derive the useful fact that $\int\!p_sdx_s\!=\!0$. As a consequence, the action associated with an outbreak in the white-noise limit is simply
\begin{equation}
S=\int\!p_{i}dx_{i}.
\label{eq:S_only}
\end{equation}
\subsection{\label{sec:PhaseSpace} Phase-space trajectories for outbreaks}
In order to compute the outbreak probabilities we need to solve Hamilton's equations and substitute the resulting trajectories into Eq.(\ref{eq:S_only}). To do so, we must understand the phase-space structure of outbreak paths. 
First we recall that in the mean-field system Eq.(\ref{MFdynamics}), the outbreak dynamics follow a heteroclinic trajectory, which starts at $t\!=\!0$ at a fixed point  $(x_{s}\!=\!1,x_{i}\!=\!0)$ and ends at the final state $(x_{s}\!=\!x_{0},x_{i}\!=\!0)$ as $t\!\rightarrow\!\infty$. In our Hamiltonian system this corresponds to a special trajectory with $p_{s}\!=0\!$ and $p_{i}\!=\!0$, or in phase space $(x_{s},x_{i},p_{s},p_{i})\!=\!(1,0,0,0)$ for $t\!=\!0$. However, in general there are an {\it infinite} number of related $x_{i}\!=\!0$ initial conditions with non-zero momenta, which one can find by solving $\dot{x}_{s}\!=\!0$, $\dot{x}_{i}\!=\!0$, $\dot{p}_{s}\!=\!0$, and $\dot{p}_{i}\!=\!0$, given 
$x_{s}\!=\!1$ and $x_{i}\!=\!0$. It is straightforward to show that the general fixed-point initial conditions are 
\begin{equation}
(x_{s},x_{i},p_{s},p_{i})_{_{\scaleto{t=0}{5pt}}}\;=\;(1, 0, \delta(R_{0}-1)/R_{0},\delta), 
\end{equation} 
where $\delta\equiv p_{i}(t\!=\!0) $ is a free parameter. 

\begin{figure}[h]
\center{\includegraphics[scale=0.201]{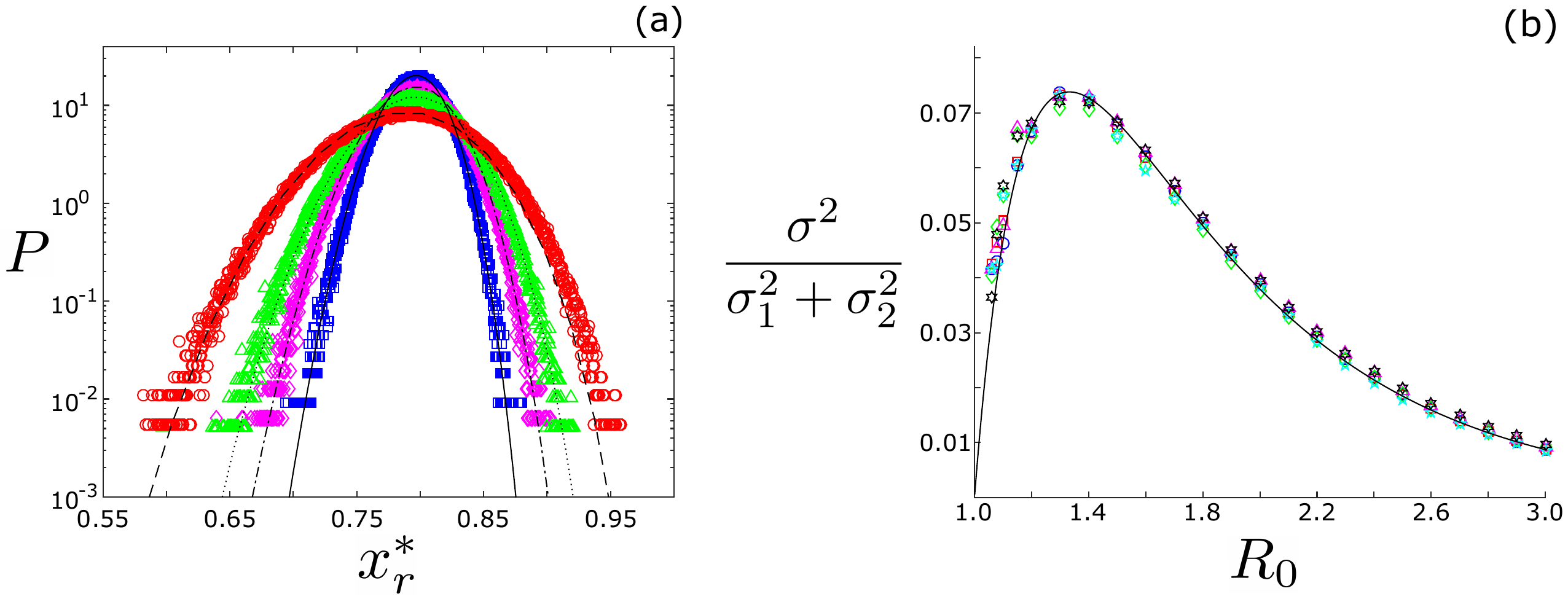}}
\vspace{-7mm}
\caption{Outbreak statistics for white noise. (a) The simulated white-noise PDFs from Eqs.(\ref{white}) for $\sigma_{1}=0.1$ and $f=0.1,\;1,\;2$ and $5$ (from narrowest to widest) with $\beta_{0}\!=\!2$. White-noise predictions for each combination are shown with curves overlaying the simulation results. (b) Variance of the outbreak PDF (normalized by noise variance) for white reaction-rate noise versus $R_0\!=\!\beta_{0}/\gamma_{0}$. The noise combinations are: blue circles ($\sigma_{1}\!=\!0.1$, $f\!=\!0.1$), red squares ($\sigma_{1}\!=\!0.1$, $f\!=\!1$), green diamonds ($\sigma_{1}\!=\!0.1$, $f\!=\!5$), magenta triangles ($\sigma_{1}\!=\!0.2$, $f\!=\!0.1$), and black hexagrams ($\sigma_{1}\!=\!0.04$, $f\!=\!1$). The black curve shows the white-noise predictions. In both panels $\gamma_{0}\!=\!1$.}
\label{fig3}
\end{figure}

As pointed out in \cite{hindes2022outbreak} for the case of demographic noise, if we propagate each of the possible initial conditions forward in time, they tend to unique final outbreak values; namely, one $x_{s}^{*}$ ($x_{r}^{*}$) for each $\delta$. Examples are shown in Fig.\ref{fig3} (b), where we plot the outbreak sizes as a function of $\delta$ for three different values of $R_{0}$. A simple algorithm for generating the outbreak distribution numerically for a fixed value of $R_{0}$ is to: (1) pick a $\delta$, (2) propagate forward with Hamilton’s equation given Eq.(\ref{Hamiltonian}) (assuming some small perturbation from the chosen fixed point), (3) compute the integral in Eq.(\ref{eq:S_only}) from the resulting trajectory, and (4) repeat for another value of $\delta$. Each $\delta$ results in a unique $x_{s}^{*}$ and $S(x_{s}^{*})$. 

The slopes of the lines in Fig.\ref{fig2}, were computed in just this way, and correspond to numerical solutions for the outbreak paths and associated $S(x_{s}^{*})$ for the chosen values of $x_{r}^{*}=1\!-\!x_{s}^{*}$. Similarly, by sweeping over values of $\delta$ we can compute the full white-noise distributions for any $x_{r}^{*}$. Examples are plotted in Fig.\ref{fig1}(a) (the narrow PDF prediction) for infection-rate fluctuations, and in Fig.\ref{fig3}(a) for different combinations of infection and recovery noise. For all Figs. \ref{fig1}(a), \ref{fig2}, and \ref{fig3}(a) the  white-noise WKB theory and simulations agree well, which demonstrates the accuracy of our general approach. In fact, by combining the method presented with the results of \cite{hindes2022outbreak}, we have a complete algorithmic solution for generating the outbreak PDF of the SIR model with general and multi-component white noise, which we return to in Sec.\ref{sec:Crossover}.



\subsection{\label{sec:WNVariance} Outbreak variance}
In the general case of noise in both $\beta$ and $\gamma$, it seems that Hamilton's equations cannot be solved analytically in a simple manner -- apart from constructing a power-series expansion in the initial-condition parameter $\delta$. The primary reason for this, in contrast to \cite{hindes2022outbreak}, is that for the reaction-rate noise discussed in this work there is {\it no conservation of momentum}. Therefore, we proceed to first calculate the variance of the outbreak-size distribution, which is related to the lowest order contribution to Eq(\ref{eq:S_only}) in $\delta$. A complete solution for the case of recovery-only fluctuations, at all orders, is given in App. \ref{sec:RecoveryOnly}. 

For the variance calculation, we attempt to find the action in the vicinity of the mean-field final outbreak fraction $1-x_0$. First, let us assume $p_i\ll 1$, to be verified a-posteriori. Equating $H=0$, yields: $p_s=(1-1/(R_0 x_s))p_i$, i.e., $p_s\ll 1$ as well. Second, we show that $p_i(t)$ remains small during the entire epidemic duration as long as the initial momentum $\delta$ is small. Writing down the Hamilton's equation for $\dot{p}_i=-\partial_{x_i}H$, and using Eq.~(\ref{Hamiltonian}) we have:
$\dot{p}_i=-(1/2)(1+f)x_i(t)p_i^2$. The solution of this differential equation is  
\begin{equation}\label{pisol}
p_i(t)\simeq\delta\left\{1+[(1+f)/(2R_0)] \ln(x_s)\delta \right\},
\end{equation}
where $p_i(0)=\delta$ is the initial condition, and we have used the fact that in the leading order in $\delta\ll 1$, 
$x_r=\int x_idt=\ln(x_s)/R_0$. 
This is legitimate as the action will have a $\delta^2$ dependence, see below, so we can substitute in ${\cal O}(\delta^2)$ terms their mean-field ${\cal O}(\delta^0)$ approximation. 

To compute $\delta$, we can use Hamilton's equations for $\dot{x}_s$, and  $\dot{x}_r=-\dot{x}_s-\dot{x}_i$, and compute $\dot{x}_r/\dot{x}_s=dx_r/dx_s$. This yields a differential equation, which can be solved with initial conditions $x_r(t=0)=0$ and $x_s(t=0)=1$, assuming that during the epidemic duration, $p_i(t)$ is almost constant within ${\cal O}(\delta)$.  Using Eq.~(\ref{pisol}) and that when  the outbreak ends $x_r^*=1-x_s^*$,   and assuming  $x_s^*-x_0\sim {\cal O}(\delta)$ (to be confirmed a-posteriori), we find
\begin{equation}\label{pi}
\delta\simeq \frac{2(1-R_0 x_0)}{(1+f) (1-x_0) x_0 (2 - R_0(1+x_0))}(x_s^*-x_0).
\end{equation}
This confirms a-posteriori that $\delta\ll 1$, under the assumption that the final susceptible fraction is close to its mean-field counterpart, i.e. $x_s^*-x_0\ll 1$.  

Finally, to compute the integral in Eq.~(\ref{eq:S_only}), it is more convenient to change variables to $x_s$, see Eq.~(\ref{pisol}). Thus, we write: $\int p_idx_i=\int p_i (dx_i/dx_s)dx_s$. Here, the Jacobian can be found  using the Hamilton's equations: $dx_i/dx_s\!=\!1/(R_0 x_s)\!-\!1\!-\![(1\!+\!f)/(R_0x_s)(1\!-\!x_s\!+\!\ln x_s/R_0)]\delta$, where again we have used mean-field results for the ${\cal O}(\delta)$ terms, namely $x_i=1-x_s+\ln(x_s)/R_0$. Putting it all together, and using Eqs.~(\ref{pisol}) and (\ref{pi}), we can perform the integration in Eq.~(\ref{eq:S_only}) over $x_s$ from $1$ to $x_s^*$, which yields the action, in the leading order in $x_s^*-x_0\sim\delta$:
\begin{eqnarray}
\label{SigmaW}
S&=&\frac{(x_s^*-x_0)^2}{2v^2}+\mathcal{O}\big((x_s^*-x_0)^3\big), \nonumber\\
v^2&=&\frac{(1+f)(x_0-1) R_0 x_0^2 [2 - R_0(1+x_0)]}{2(1-R_0 x_0)^2}.
\end{eqnarray}
Indeed, having obtained a $\delta^2$ dependence of the action corroborated our assumptions a-posteriori.  
Here, $v\!=\!v(R_0)$ is the (rescaled) variance of the outbreak-size distribution. 
Namely, remembering that we have sought the outbreak-size PDF as $P(x_s^{*})\!\sim\!\exp\!\left[-S(x_s^{*})/\sigma_{1}^2\right]$, the variance of the outbreak-size distribution in the limit of white reaction-rate noise, $\sigma_{w}^{2}$, is
\begin{eqnarray}
\label{VarW}
\sigma_{w}^2=\sigma_{1}^2 v^2. 
\end{eqnarray}
%
%


We can test our predictions for the outbreak variance in the white-noise regime by performing stochastic simulations of Eqs.(\ref{white}) with different values of $R_{0}$ and different combinations of noise. Results are shown in Fig.~\ref{fig3}(b). First, one can see an interesting behavior where the variance receives a maximum at $R_0\simeq 1.33$, similar to the adiabatic regime shown in Fig.\ref{fig1} (b). Here, however, the maximum variance occurs for an $R_{0}$ that is independent of the noise amplitude and noise combination, unlike adiabatic noise. The reason for the maximum appearing around $1.33$ is that for this value of $R_0$ the mean outbreak fraction is approximately obtained at $x_r^*\simeq 0.5$ which maximizes the variance possibility. In addition, we note that the predicted outbreak variance in the white-noise regime only depends on $R_{0}$ and the total variance of the reaction-rate noise, $\sigma_{1}^{2}+\sigma_{2}^{2}$. For example, in Fig.~\ref{fig3}(b) we show the predicted scaling collapse to a single function of $R_{0}$ of the simulated outbreak variance resulting from different combinations of noise. In general, we would expect infection-rate and recovery-rate noise to produce additive variance (since the two noise sources are independent), but the fact that their prefactor dependence on $R_{0}$ is identical is interesting. On the other hand, one can check that this symmetry between infection and recovery noise disappears for higher-order statistics, e.g., by repeating the above calculation to $\mathcal{O}(\delta^{3})$ for the third central moment. 

\section{\label{sec:Crossover} CROSSOVER WITH CORRELATION TIME AND SYSTEM SIZE}

Now that we have analyzed the outbreak-distribution in limiting cases (including App. \ref{sec:RecoveryOnly}), we next address when the various limiting regimes apply. In particular, we examine the cross-over behavior of the stochastic SIR model as a function of the reaction-rate noise correlation time  and  population size; the latter has been assumed infinite so far. We use as our metric the variance of the outbreak-size distribution since it is the lowest order statistic not captured by mean-field theory.  

\begin{figure}
\center{\includegraphics[scale=0.215]{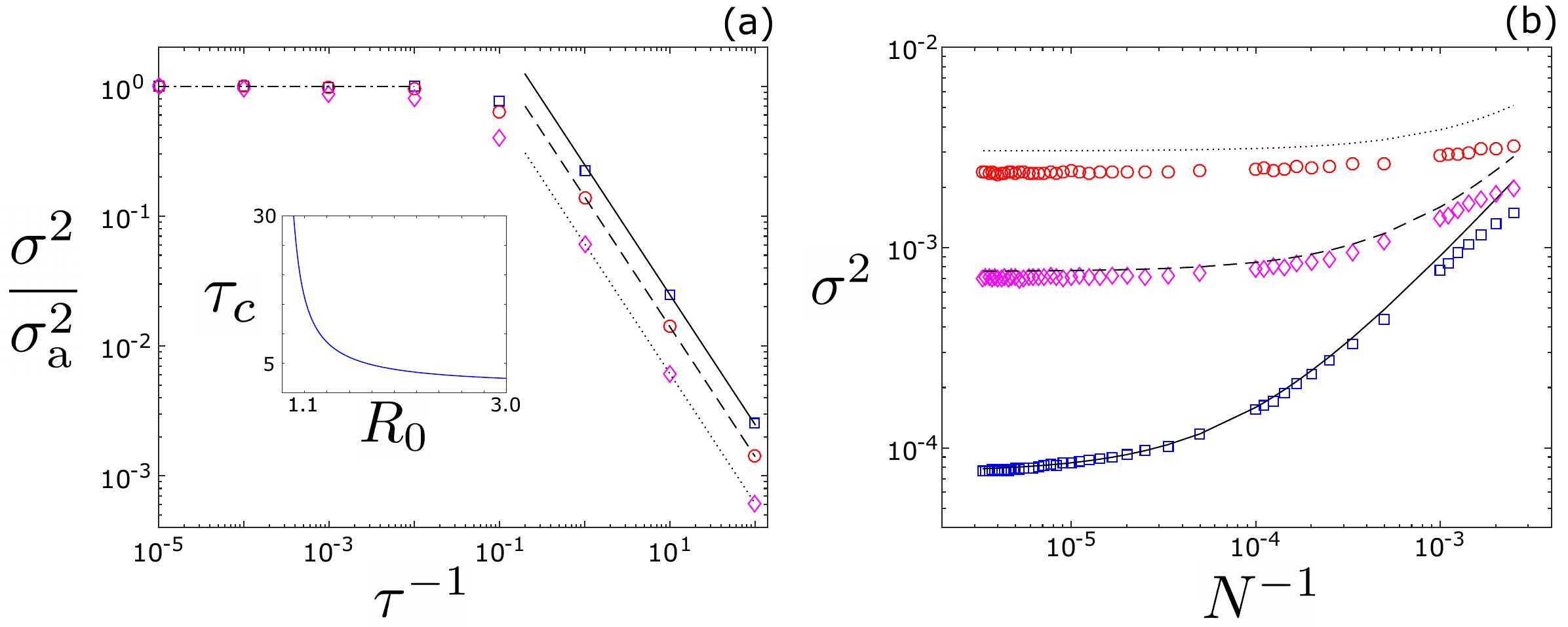}}
\vspace{-7mm}
\caption{Variance of the final outbreak size versus the inverse of the correlation time (left) and $N$ (right). Panel (a): The variance (rescaled by the predicted adiabatic-limit) as function of $\tau^{-1}$ for $\sigma_{\beta}=0.04\beta_{0}$ and $\beta_{0}=2.0$,  $1.5$ and  $1.2$ (top to bottom). The lines are the white-noise predictions. The inset shows the crossover time, Eq.(\ref{eq:CrossoverTau}) versus $R_{0}$. Panel (b): The variance versus $N^{-1}$ for $\tau=10$, $1$ and $0.1$ (top to bottom) with $\beta_{0}=2.0$ and $\sigma_{\beta}=0.1\beta_{0}$. The lines are the white-noise predictions, which are the sum of the variances from reaction-rate and demographic noise. For both panels $\gamma_{0}\!=\!1$. 
}
\label{fig4}
\end{figure}

First, we remain in the $N\!\rightarrow\!\infty$ limit, and try to understand 
how small (large) $\tau$ has to be in order to produce effectively white (adiabatic) outbreak statistics. To do so, we plot  in Fig.\ref{fig4}(a) the variance of the outbreak-size PDF found from simulating Eqs.(\ref{Langeqs}) versus the (inverse) correlation time  $\tau$ for three values of $R_{0}$ and fixed $\sigma_{\beta}$. Note that the outbreak variance is normalized by the adiabatic limit, Eq.(\ref{SigmaAdi}), so that each simulation series approaches unity for small $\tau^{-1}$. In addition to the adiabatic limit, for comparison we plot predictions for white-noise, Eq.(\ref{SigmaW}), with lines. In the latter case, the $\tau$-dependence comes from the definition of the white-noise variance, Eqs.(\ref{whiteNoiseAmplitudes}).

Figure \ref{fig4}(a) has several important features. First, we point out that the outbreak variance has a maximum in the adiabatic limit, meaning that for fixed infection-rate noise variance, the SIR model dynamics is most sensitive to slow noise. This effect is observed in other population models as well\cite{PhysRevLett.111.238101,PhysRevLett.125.048105}. For the SIR model, the primary reason is that even relatively small fluctuations in $\beta$ can bring an epidemic closer to the $R_{0}\!=\!1$ threshold. If the noise is slowly varying, in particular, the effect is felt over the full time-course of the epidemic wave, resulting in potentially much smaller outbreak than the mean-field. As mentioned in Sec.\ref{sec:LimitAdiabatic}, this produces highly skewed PDFs with significant probabilities for small outbreaks, and hence large outbreak variance.  In contrast, in Fig.\ref{fig4}(a) we can see that the white-noise predictions (the lines) are accurate for quite large values of $\tau$. In fact, for each value of $R_{0}\!=\! 2,\;1.5,$ and $1.2$ (from top to bottom), we can see that the white noise prediction remains valid for correlation times on the order of the recovery time,  $\tau\!\sim\!\gamma_{0}^{-1}\!=\!1$.

In general, the crossover point in $\tau$ between white- and adiabatic-noise regimes has some $R_{0}$ dependence: the smaller $R_{0}$, the larger $\tau$ can be for the white-noise results to be valid, since effectively as the epidemic gets closer to threshold the dynamics slows down, making even slowly-varying noise potentially fast.
An estimate for the crossover time, $\tau_{c}$, can be found by solving $\sigma_{a}^{2}\!=\!\sigma_{w}^{2}$, or
\begin{eqnarray}
\label{eq:CrossoverTau}
\tau_{c}=\dfrac{R_{0}^{2}}{2v^{2}}\dfrac{\sigma_{a}^{2}}{\sigma_{\beta}^{2}}=\frac{R_0(x_0-1)}{(1+f)(2-R_0(1+x_0)}, 
\end{eqnarray}
where we have used Eqs.~(\ref{SigmaAdi}) and (\ref{SigmaW}), valid for small noise. Evidently, $\tau_c$ depends only on $R_0$ and not, e.g., on the noise variance for small noise. The crossover time is plotted in the inset of Fig.\ref{fig4}(a), which for the typical model parameters of $R_{0}\!-\!1\sim\mathcal{O}(1)$ remains near the recovery time scale (or unity in our chosen units).

Now that we have an estimate for cross-over times, we can situate the inferred RSV contact-rate fluctuations and determine what regime they fall into. By plugging in the median and quartile inferred parameter values given in Sec.\ref{sec:RSVModelFit} into Eq.(\ref{eq:CrossoverTau}), and using Eqs.~(\ref{SigmaAdi}) and (\ref{SigmaW}), we find that the ratio of the noise correlation time to the cross-over time, $\tau/\tau_{c}$, falls between $0.1\!-\!0.2$. As the $\tau$ estimates are substantially smaller than the cross-over times, we expect the outbreak-size statistics to be well approximated by the white-noise theory. Hence, our analytical results can be used to make quantitative estimates for future RSV outbreak size probabilities, assuming parameters remain relatively similar to the 2019-2020 epidemic.       
 
The second crossover that we consider is that of finite system size. Namely, how large does a population have to be before demographic noise becomes irrelevant compared to reaction-rate noise? For this exploration we perform a discrete time stochastic simulation (with small time steps) of the discrete state reactions defined for the SIR model in Sec.\ref{sec:SIRModel} (above Eqs.\ref{MFdynamics}) while  the reaction-rates fluctuate according to the OU processes in Eqs.(\ref{Langeqs}). In Fig.\ref{fig4}(b) we plot the outbreak-size variance as a function of $N^{-1}$ for three values of the correlation time $\tau=10,\;1,$ and $0.1$ (from top to bottom). The curves are the expected total white-noise variance, $\sigma^{2}_{w,tot}$, which is a sum of reaction-rate and demographic noise 
\begin{eqnarray}
\label{eq:WhiteTotal}
&\sigma^{2}_{w,tot}=\dfrac{(\sigma_{1}^{2}+\sigma_{2}^{2})(x_0-1) R_0 x_0^2 [2 - R_0(1+x_0)]}{2(1-R_0 x_0)^2}\;\;+\nonumber \\
&\dfrac{x_{0}(1-x_{0})(R_{0}^{2}x_{0}+1)}{N(R_{0}x_{0}-1)^{2}}.
\end{eqnarray}
The demographic piece was calculated in \cite{hindes2022outbreak}. Note that here we have assumed that the total variance is the sum of the variances from the independent noise sources\cite{IntrinsicNoiseInGeneRegulatoryNetworks}, and that this result holds for  $\sigma_{1}^{2},\sigma_{2}^{2},N^{-1}\!\ll\! 1$.

In Fig.\ref{fig4}(b) we can see that for large system sizes the variance becomes flat with respect to changes in $N$ and approaches approximately the expected white-noise limit, Eq.(\ref{SigmaW})  -- especially for the two smaller values of $\tau$ where the white-noise approximation is more appropriate. On the other hand, the crossover can occur for quite large system sizes, e.g., $N\!\sim\!10^{5}$ for $\tau\!=\!0.1$ and $N\!\sim\!10^{4}$ for $\tau\!=\!1$, meaning that demographic noise tends to persist if the reaction-rate noise is fast, but disappears quickly with $N$ if the noise is slow; notice that the top series with $\tau\!=\!10$ has almost no $N$-dependence. 

\section{\label{sec:Conclusions} CONCLUSIONS}

Temporal fluctuations in the parameters that control contagion dynamics
are inevitable, and have been shown in many epidemic data analyses. 
Motivated by this, we analyzed the effects of fluctuating infection and recovery rates on 
the outbreak-size distribution in the canonical SIR model. The SIR reaction rates were modeled with Ornstein Uhlenbeck noise, allowing us to extract the outbreak statistics as a function of the noise standard deviations and correlation times. Our simple choice was demonstrated by performing a model inference of the 2019-2020 RSV season in the US, where we observed significant temporal fluctuations in infectious contact rates.  

In terms of analytical results, we found solutions for the outbreak-size distribution in the adiabatic and white noise regimes, and showed that the distributions can be highly skewed with significant probabilities for large fluctuations away from mean-field predictions. Interestingly, we discovered that the outbreak variance is generally maximized for a value of the basic reproductive number that depends on the correlation time of the noise, which in the white-noise limit is independent of where noise resides (infection or recovery). In addition, we compared the typical fluctuations emerging from demographic and reaction-rate noise and determined the population sizes, correlation times, and reproductive numbers that places noisy SIR systems in the various limiting regimes. Altogether then, our work illustrated a rich interplay between noise and outbreak dynamics -- depending sensitively on fundamental noise characteristics and population size. 

Currently the theory presented pertains to well-mixed populations in which individuals come into contact with a contagion with homogeneous rates. In actuality the contact rates within a population can be highly heterogeneous and/or spatially distributed, and therefore, an important extension of our work will include the generalization of the outbreak-size distribution to complex network topology. Another common assumption that we used, which is only an approximation, was the implicit exponential waiting times for both the infection and recovery processes. Future work should include generalization to gamma and other more realistic waiting-time distributions. Finally, our work has relied substantially on small-noise assumptions allowing us to focus on the dominant, exponential contribution to the outbreak-size distribution. Corrections to this approach, which would include next-order contributions for larger noise amplitudes, are an important avenue for future analysis.     

\section{\label{sec:Acknowledgements} ACKNOWLEDGMENTS}
JH and IBS were supported by the U.S. Naval Research Laboratory funding
(N0001419WX00055), and the Office of Naval Research (N0001419WX01166) and
(N0001419WX01322). MA was supported by the Israel Science Foundation Grant No. 531/20. 

\section{\label{sec:Appendices} APPENDICES}
\appendix
\section{\label{sec:RSV_DataAnalysis} RSV data analysis}
We implement a time-discrete version of an SIR model to carry out Bayesian parameter inference on the hospitalization data for the 2019-2020 season of respiratory syncytial virus (RSV) in the US\cite{CDCdata}. At day $d$, the daily model updates follow the dynamics:
\begin{subequations}
\begin{align}
i_d &= \beta_d \cdot S_{d-1} \cdot I_{d-1},\\
S_{d} &= S_{d-1} - i_{d-1},\\
I_{d} &= I_{d-1} + i_{d} - \gamma \cdot I_{d-1}\\
R_{d} &= R_{d-1} + \gamma \cdot I_{d-1}
\end{align}
\end{subequations}
where $(S_d, I_d, R_d)$ are the susceptible, infected and recovered fractions ($S_d + I_d + R_d = 1$), respectively, and $i_d$ is the fractional incidence. The parameters $\beta_d$ and $\gamma$ refer to the contact and recovery rates, respectively.

In our inference model, the dynamics of $\beta_d$ are given by a daily-discretized version of the Ornstein-Uhlenbeck process:
\begin{eqnarray}
\beta_1 &\!=\!& \mathcal{N}(\beta_0, \sigma_\beta),\\
\beta_d &\!=\!& \beta_{d - 1} \!-\! \alpha  (\beta_{d - 1} \!-\! \beta_0) \!+\!  \mathcal{N}(0, 2 \alpha  \sigma_\beta), \;\;\; d \!>\!1\!,
\end{eqnarray}
where $\mathcal{N}(\mu, \sigma)$ is a normal random variable with mean $\mu$ and variance $\sigma^2$. 

The parameter inference process is done by tying the daily-discretized statistical model to the number of daily hospitalizations contained in the data $H_d$ by the Poisson observation process
\begin{align}
H_{d} = \text{Pois}(\eta \cdot N \cdot i_d)
\end{align}
where $\eta$ is the hospitalization rate and $N$ denotes the total population. With this model we do Bayesian parameter inference using the platform {\tt Stan} via the {\tt R} package {\tt rstan}\cite{Rstan,STANexampleCOVID}; {\tt R} code for the model inference is available upon request.

\section{\label{sec:RecoveryOnly} Action for white-noise fluctuations in recovery}

In addition to small fluctuations in the outbreaks, we can gain further analytical insight into the outbreak distribution for white noise by looking at other limiting-case scenarios. One such scenario is when the recovery-noise dominates over infection noise, namely $f\!\gg\!1$ while $\sigma_{\gamma}\!\ll\!1$. In this limit, the Hamiltonian reduces to
\begin{equation}
H=x_i[p_i(R_0 x_s-1)-R_0 x_s  p_s]+\frac{1}{2}fx_i^2p_i^2. 
\label{eq:H_recovery}
\end{equation}
As in Sec.\ref{sec:WNVariance}, given the simpler Hamiltonian~(\ref{eq:H_recovery}), the $H\!=\!0$ condition can be combined with Hamilton's equations $\dot{x}_{s}$ and $\dot{p}_{i}$, to give $p_{i}$ as an explicit function of $x_{s}$ and the initial condition for $\delta\equiv p_{i}(t\!=\!0)$:
\begin{equation}
p_{i}(x_{s})=\frac{2\delta}{2-f\delta\ln(x_{s})/R_{0}}.
\label{eq:pi_recovery}
\end{equation}
Similarly, if we substitute Eq.~(\ref{eq:pi_recovery}) and the zero-energy condition into $\dot{x}_{i}$
and divide by $\dot{x}_{s}$, we get the following differential equation for $x_{i}$ as a function of $x_{s}$: 
\begin{equation}
\frac{dx_{i}}{dx_{s}}=-1 +\frac{1}{R_{0}x_{s}} -\frac{2fx_{i}\delta}{R_{0}x_{s}\Big(2-f\delta\ln(x_{s})/R_{0}\Big)}. 
\label{eq:dxi_recovery}
\end{equation}
Next, we can solve for $x_{i}(x_{s})$ by separating the fraction of the population infected into a product of two functions that depend on $x_{s}$, i.e., $x_{i}(x_{s})\!=\!u(x_{s})v(x_{s})$. By substituting the product form into Eq.(\ref{eq:dxi_recovery}), setting $u\;dv/dx_{s}\!=\!-1 +1/(R_{0}x_{s})$, and conditioning $u(x_{s}\!=\!1)\!=\!1$ and $v(x_{s}\!=\!1)\!=\!0$, we find:   
\begin{equation}
\label{eq:uv_recovery}
u(x_{s})\!=\! \frac{1}{4}\Bigg(\!\!\frac{f\delta}{R_{0}}\ln(x_{s})-2\!\Bigg)^{\!\!2}\!\!,\;\; 
v(x_{s})\!=\!\int_{1}^{x_{s}}\!\!\!\dfrac{1-R_{0}x_{s}'}{R_{0}x_{s}'u(x_{s}')}\;dx_{s'}, 
\end{equation}
where the integral in Eq.(\ref{eq:uv_recovery}) can be expressed in terms of incomplete exponential integrals, though the formula is cumbersome. As $x_{i}(t\!\rightarrow\!\infty)\!\rightarrow\!0$, Eq.~(\ref{eq:uv_recovery}) gives us a condition for the final outbreak $1-x_{s}^{*}$. Namely, given $R_{0}$ and $\delta$, we can solve $v(x_{s}^{*})=0$
for the unique value of $x_{s}^{*}$.

The final step for calculating the action associated with a given outbreak in the limit of recovery-only fluctuations, is to differentiate $x_{i}(x_{s})\!=\!u(x_{s})v(x_{s})$ and substitute Eqs.(\ref{eq:pi_recovery}) and (\ref{eq:uv_recovery}) into Eq.(\ref{eq:S_only}). The result is the following limiting-case action
\begin{equation}
\label{eq:S_recovery}
S=\bigintss_{1}^{x_{s}^{*}}\!\!\dfrac{2\delta\Big(-1+\dfrac{1}{R_{0}x_{s}}\Big)}{\Big(f\delta\ln(x_{s})/{R_{0}}-2\Big)}\;dx_{s},  
\end{equation}
where as mentioned, the boundary condition for the integral (the final outbreak size) can be determined numerically from the condition $v(x_{s}^{*})=0$. 

\bibliography{ExtremeOutbreaks,bibtex_IBS}

\end{document}